\documentclass[9pt,twocolumn,twoside]{pnas-new}

\templatetype{pnasresearcharticle} 

\usepackage{chemmacros}  
\usepackage{xcolor}
\usepackage{upgreek}    
\usepackage{gensymb}
\usepackage{lineno}
\linenumbers

\begin{document}

\title{Non-additivity in many-body interactions between membrane-deforming spheres increases disorder}


\author[a]{Ali Azadbakht}
\author[b]{Thomas R. Weikl}
\author[a,1]{Daniela J. Kraft}

\affil[a]{Soft Matter Physics, Huygens-Kamerlingh Onnes Laboratory, Leiden University\\
PO Box 9504, 2300 RA Leiden, the Netherlands}
\affil[b]{Department of Biomolecular Systems, Max Planck Institute of Colloids and Interfaces\\
 Am M\"uhlenberg 1, 14476 Potsdam,Germany}

\leadauthor{Azadbakht}

\significancestatement{
Membrane proteins can cause distortions in the lipid bilayer, leading to interactions based on the imposed curvature. However, these interactions are not simply additive, making it challenging to derive the forces between multiple membrane-deforming objects based solely on their pairwise interactions. Furthermore, the equations governing these interactions are complex, resulting in limited and often contradictory theoretical and numerical predictions. In this manuscript, we introduce a new model system that allows quantitative observation of the interactions between many membrane-deforming micrometer-sized spheres. We find that they form hexagonal clusters, which become more disordered as the number of spheres increases.
}


\authorcontributions{A.A. and D.J.K. designed the research; A.A. performed the experimental research; T.R.W. designed and performed the numerical modeling; all authors analyzed the data and wrote the paper.}
\authordeclaration{The authors declare no competing interest.}

\correspondingauthor{\textsuperscript{1} To whom correspondence may be addressed. kraft@physics.leidenuniv.nl}


\begin{abstract}
Membrane-induced interactions have been predicted to be important for the organization of membrane proteins. Measurements of the interactions between two and three membrane deforming objects have revealed their non-additive nature. They are thought to lead to complex many-body effects, however, experimental evidence is lacking to date. We here present an experimental method to measure many-body effects in membrane-mediated interactions using colloidal spheres placed between a deflated giant unilamellar vesicles and a planar substrate. The thus confined colloidal particles cause a large deformation of the membrane while not being physochemically attached to it and interact through it. Two particles are found to attract with a maximum force of 0.2~pN. For three particles, we observe a preference for forming compact equilateral triangles over a linear arrangement. We use numerical energy minimization to establish that the attraction stems from a reduction in the membrane-deformation energy caused by the particles.  
Confining up to 36 particles, we find a preference for hexagonally close packed clusters. However, with increasing number of particles the order of the confined particles decreases, while at the same time, diffusivity of the particles increases. Our experiments for the first time show that the non-additive nature of membrane-mediated interactions affects the interactions and arrangements and ultimately leads to spherical aggregates with liquid-like order of potential importance for cellular processes. 

\end{abstract}

\dates{This manuscript was compiled on \today}
\doi{\url{www.pnas.org/cgi/doi/10.1073/pnas.XXXXXXXXXX}}

\maketitle
\ifthenelse{\boolean{shortarticle}}{\ifthenelse{\boolean{singlecolumn}}{\abscontentformatted}{\abscontent}}{}

\firstpage[8]{3}

\dropcap{M}any cellular processes rely on the cooperation of multiple membrane proteins. One contributing factor to their effective self-assembly is thought to arise from the membrane deformations that they induce. While predicted by theory almost 30 years ago\cite{Goulian1993}, these membrane-mediated interactions have only recently been unequivocally measured in colloidal model systems. For pairs of spherical colloidal particles adhered to and thereby deforming giant unilamellar vesicles, the interaction was found to be attractive over several particle diameters and its strength varied depending on the adhesion method.\cite{VanDerWel2016,Sarfati2016} 

However, these membrane-mediated interactions have been predicted to be non-additive \cite{Saric2013,Yolcu2014a,Weikl2018,Galatola2023}, and cooperativity effects observed in simulations demonstrate that the knowledge of the interaction energy between two spheres is not sufficient to understand the behavior of many such interacting particles \cite{Saric2012,Laradji2020}. 
Indeed, for three spheres fully wrapped by the lipid membrane, a recent experimental study from some of us has unveiled that  the presence of a third particle did not enhance the strength of the interactions but rather increased the preferred distance between the other two particles~\cite{Azadbakht2023BPJ}. In addition, two distinct preferred states for three membrane-deforming spheres were found: a linear arrangement and an equilateral triangular arrangement, where the latter is preferred for slightly larger distances between the particles.\cite{Azadbakht2023BPJ} These observations suggest that arrangements of many particles might be hexagonally ordered, in line with early experiments on surfactant membranes\cite{Ramos1999} and qualitative observations on lipid membranes\cite{Koltover1999}, and coarse-grained simulations of Janus particles bound to planar membranes~\cite{Zhu2023}, and that many-body effects play an important role. However, such  many-body effects due to non-additive interactions have not been  explored in any quantitative or systematic way in experiments.   

Experimental models have so far been hampered by various challenges when studying the behavior of multiple membrane-deforming objects. Previous studies investigated membrane deformations using colloidal particles that adhere to giant unilamellar vesicles (GUVs) through electrostatic charges~\cite{Ramos1999}, depletion forces~\cite{Spanke2020}, or strong ligand-receptor bonds~\cite{VanDerWel2016,Sarfati2016,Koltover1999,Azadbakht2023NL}. However, electrical charges and depletion effects may interfere with measurements of membrane-mediated forces as they also act between the particles in the absence of the membrane. Ligand-receptor based adhesion may lead to non-uniform deformations if the ligands are not distributed homogeneously on the particle surface and, if sufficiently strong, imply an irreversible attachment of the membrane to the particle, thereby limiting optimization of the membrane shape to lower the bending energy. Moreover, accurate extraction of the interaction potential becomes a challenge when multiple particles on a GUV are involved, since imaging and tracking requires the particles to be located in the same detection plane, especially when increasingly extensive amounts of data need to be collected~\cite{VanDerWel2016, Azadbakht2023BPJ}. 

To address these issues, we here present a novel attachment-free model system that enables simple and rapid quantification of membrane-deformation-induced interactions among colloidal spheres. Using optical traps in combination with confocal microscopy, we position colloids beneath a deflated GUV placed on top of a flat substrate. The presence of the colloidal spheres causes significant deformations of the GUV while allowing for membrane remodeling due to the absence of an attachment to the membrane. Because the particles remain in the same imaging plane,  two-, three-, four-, and many-particle arrangements can be measured with high accuracy. We find that the particles organize in compact hexagonally closed packed clusters which become progressively more disordered with increasing number of particles. The forces measured between two and three particles are in quantitative agreement with forces determined from numerical minimization of the membrane bending energy at constrained volume of the solvent pockets in which the particles are confined. The minimization results indicate that the increase in disorder with increasing number of particles in clusters can be understood from a coalescence of these solvent pockets, which allows for longer-ranged, energetically less costly membrane deformations around larger particle clusters.
Together, our combined experimental and numerical study quantitatively demonstrates the surprising effects of non-additivity on the interactions and arrangements of membrane-deforming particles. 

\section*{Results and Discussion}
\begin{figure*}
\centering
\includegraphics[width=1\textwidth]{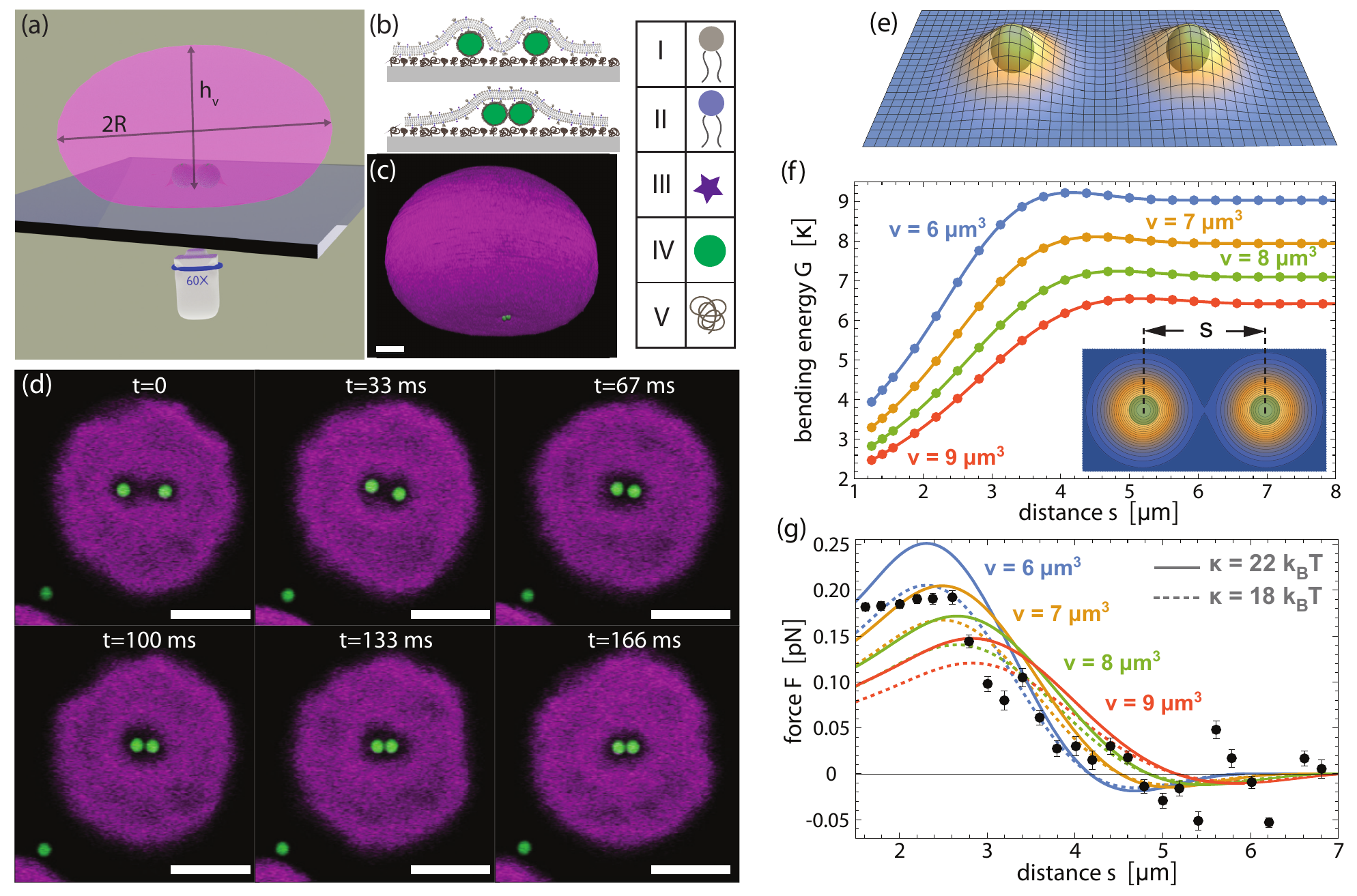}
\caption{\label{fig:fig1}\textbf{Experimental setup and membrane-mediated force between two spheres.}
(a) Schematic depicting the experimental approach in which we pull colloidal particles between a sessile vesicle and a microscope slide using optical tweezers. Parameters height $h_v$ and diameter $2R$ to describe the vesicle shape are indicated.   
(b) The particles and associated solvent initially deform the membrane individually (top panel) and cause a long-ranged deformation of the membrane. 
Upon approach (bottom panel) the membrane shape can remodel to bring two particles in contact, which costs less bending energy. Detailed schematic of the system where I-DOPC, II-DOPE, III-Rhodamine B, IV-Polystyrene colloid, V-PEG (not to scale).
(c) 3D reconstruction of a confocal image stack showing the GUV with two membrane-deforming particles. 
(d) Snapshot image sequence of two spheres approaching each other after release of the optical traps due to the interaction induced by their membrane deformation; Scale bars are 5~$\mu$m.
(e) Calculated minimum-energy membrane shape for two spheres  at distance $s = 5$ $\upmu$m confined in pockets of volume $v = 7$ $\upmu{\rm m}^3$ per sphere. 
(f) Bending energy $G$ versus distance $s$ from energy minimization with constrained volume $v = 6$ (blue), $7$ (yellow), $8$ (green), and $9$ $\upmu{\rm m}^3$ (red) per sphere.  Lines represent 12th order polynomial fits of minimization results (points). The inset shows a contour plot of the minimum-energy shape in (e).  
(g) Comparison of measured forces (data points for a vesicle with $h_v=32$~$\mu$m and $2R=40$~$\mu$m.) and calculated forces (lines) versus particle distance for different solvent pocket volumes. The force curves correspond to derivatives of the energy curves in (f) for the membrane bending rigidities $\kappa = 18$ $k_BT$ (dashed lines) and $22$ $k_BT$ (full lines) within the range of rigidity values measured for DOPC membranes \cite{Faizi2020}. Error bars are standard error.}.
\end{figure*}

Our model system to measure many-body membrane-mediated interactions consists of 1.25~$\upmu$m fluorescently labelled colloidal polystyrene spheres (depicted in green) and model lipid membranes experimentally realized by fluorescently labelled Giant Unilamellar Vesicles (GUVs, depicted in magenta), see Figure~\ref{fig:fig1}a and c. GUVs 30-50~$\upmu$m in diameter were produced by electroswelling a mixture of 97.5\%wt $\Delta$ 9-cis 1,2-dioleoyl-sn-glycero-3-phosphocholine) (DOPC), 2\%wt DOPE-PEG2000 (1,2-dioleoyl-sn-glycero-3-phosphoethanolamine-N-[(polyethylene glycol)-2000]) to prevent adhesion of the colloidal particles, and 0.5\%wt DOPE-Rhodamine as fluorescent label (see Figure~\ref{fig:fig1} and methods for details). The surrounding medium consisted of a saline solution (80\% vol of an aqueous 310~mM glucose solution and 20\% vol of a 150~mM NaCl solution (pH~7.0)), which leads to a Debye screening length of less than 5~nm to suppress unwanted interactions by electrical charge. The inside solution of the GUVs was designed such that it had a higher density than the surrounding medium. Upon deflation of the vesicle due to an intentional osmotic pressure difference between inside (300~mOsm sucrose) and outside (308~mOsm glucose mixture), the density difference between inner and outer solution $\Delta \rho~=~22.6~{\rm kg}/{\rm m}^3$ leads to sedimentation and flattening of the GUVs onto the coverslip. As a consequence, a large area where the membrane is flat appears as shown in Figure~\ref{fig:fig1}(a).

This flattened membrane of the sessile vesicle allows us to induce deformations by colloidal particles without requiring chemical bonding between the particle and the membrane. We achieve this by dragging the colloidal spheres underneath the vesicle using optical tweezers, such that the particles are sandwiched between the deflated GUV and the cover slip, see Figure~\ref{fig:fig1} and Movie~S1. To prevent adhesion of membrane or particles, the glass substrate was coated by an acrylamide brush. Similarly, the particles were sterically stabilized by a brush of polyethylene glycol (PEG5000)~\cite{VanDerWel2017a}, and the vesicles were doped with 2\%~PEGylated lipids to decrease non-specific interactions (see Methods section for experimental details). 

Placing the colloidal particles underneath the membrane incurs an energetic cost for bending the membrane around the particle. 
In addition, the confinement of the particle to quasi 2D lowers its entropy compared to the freely dispersed state. 
With no energetic gain, this state is thus unfavorable compared to the state where the particles are freely dispersed, and the colloidal particles will therefore ultimately escape the confinement. However, this only occurs after the colloid has diffused towards the outside of the vesicle, typically taking more than 200 seconds or longer. While the particle is confined underneath the membrane, it interacts with other membrane-deforming particles, allowing us to observe in a single confocal imaging plane the interactions and arrangements of multiple colloidal particles, see Figure~\ref{fig:fig1}d.

We start by measuring the interaction between two spherical membrane-deforming particles as a reference. In contrast to our earlier work,\cite{VanDerWel2016,Azadbakht2023BPJ} here the membrane is not attached to the particle and hence can change its contact area on the particles in remodeling its shape. We measure the force $F$ between two particles for different distances $s$ by measuring the displacement of the particle from the center of the optical trap, assuming a linear force regime, see Movie~S2 and Methods section. To ensure that the laser did not leave additional indentations at the interface, the intensity of the trap laser never exceeded 1~mW. We find that the attractive force appears when particles come closer than 4-5~$\upmu$m and increases upon approach. It has a maximum of $F=0.2$~pN at about $s=2.7$~$\upmu$m and then decreases slightly to stay almost constant at about 0.17~pN for even smaller values of $s$. 
Once two particles are in touch, they do not separate spontaneously again.

To better understand this membrane-mediated interaction, we numerically determined the minimum-energy shapes of the membrane around two particles (Fig.~\ref{fig:fig1}e). In our numerical approach, the shape of the membrane is described by its local height $h$ above the substrate plane, and the non-linear bending energy of the membrane is minimized after discretization at a length scale of 100 nm, which is much smaller than the particle diameter (Methods). The spherical shape of the particles is taken into account by constraints on the height of the membrane above the particles during minimization.
Besides the particle distance, an important parameter in the minimization approach turns out to be the volume $v$ of the membrane-covered solvent pocket in which each particle is confined. To estimate this volume from experiments, we fitted the vesicle's contour around a single particle to a parabola and revolved it around the z-axis, see Figure S1 in the SI, and found it to be on average 7.8~$\pm$~1.6~$\upmu$m$^3$. Fig.~\ref{fig:fig1}f shows the membrane bending energy as a function of particle distance $s$ for different values of confinement volume $v$ per particle, which is constrained in the minimization (Methods), around this experimental estimate. For values of $v$ close to the experimental estimate and bending rigidities $\kappa$ in the range 18 to 22~$k_BT$ previously measured for DOPC membranes~\cite{Faizi2020,Nagle2013}, the force curves in Fig.~\ref{fig:fig1}g obtained from the derivatives of the energy curves in Fig.~\ref{fig:fig1}f are in good agreement with the experimentally measured forces (data points). Our minimization results indicate that the range of the interaction decreases with the confinement volume $v$ per particles, while the interaction strength increases. We have neglected the membrane tension $\sigma$, which can affect curvature-mediated interactions~\cite{Weikl1998,Simunovic2015b}, in our calculations, because this tension appears to be very low in our setup ($\sigma < 1$~nN/m,  measured from the fluctuation spectrum). For such small tension values, the characteristic length scale $\lambda=\sqrt{\kappa/\sigma}$ for the crossover from bending- to tension-dominated membrane energies exceeds 9~$\upmu$m and is, thus, larger than the particle interaction range.

\begin{figure*}
\centering
\includegraphics[width=1\textwidth]{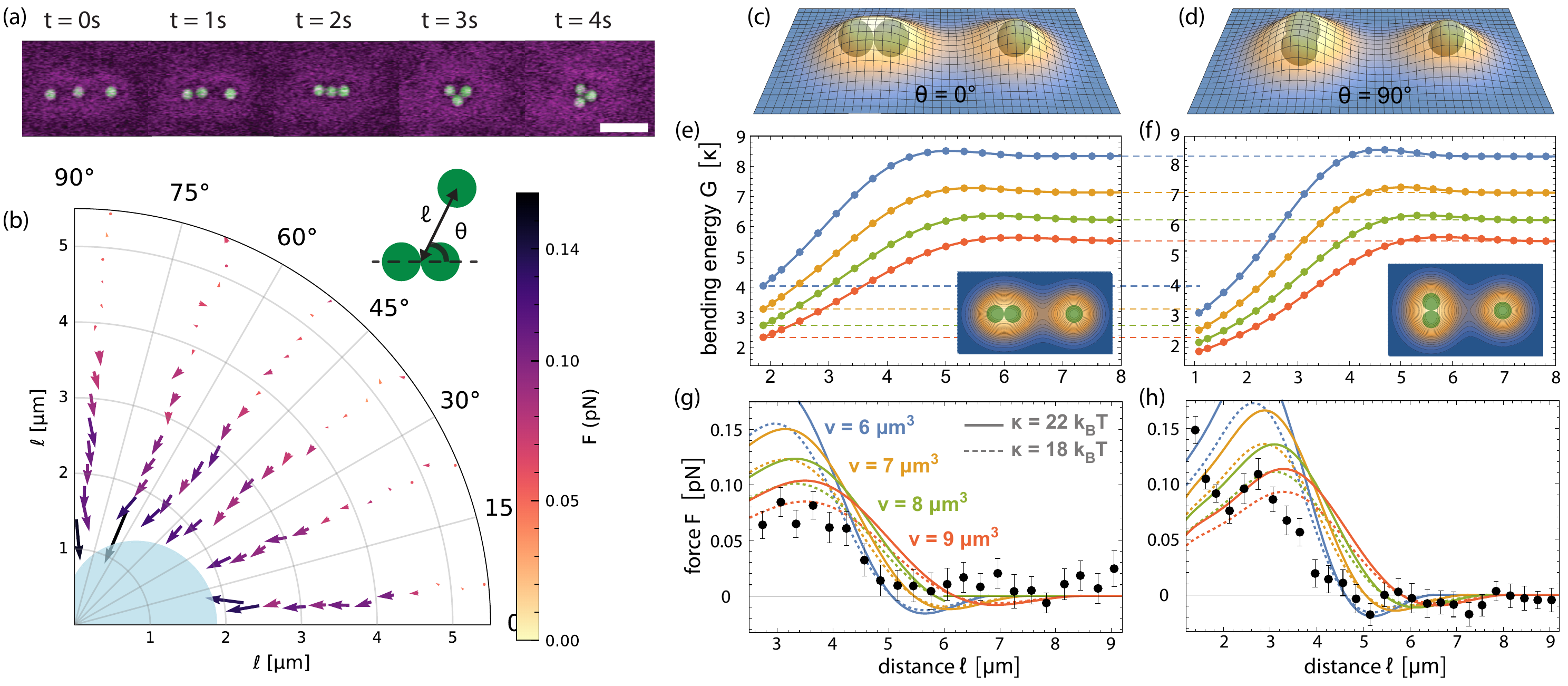}
\caption{\label{fig:three_particles}\textbf{Membrane-mediated interaction between three spheres}
(a) Snapshots in time taken from a movie (SI Movie~S3) showing the rearrangements of three spheres after releasing the optical traps: particles rearrange from a linear arrangement into an equilateral triangle; scale bars are 5~$\mu$m, vesicle with $h_v=36$~$\mu$m and $2R=52$~$\mu$m.
(b) Polar plot of the force ($\vec{F}$) acting on a sphere in the vicinity of two particles trapped in close contact by optical tweezers, where $\ell$ is the distance between the center of the sphere and the center of the mid line of the two touching spheres (illustrated in the schematic) and $\theta$ is the angle between the major axis of the two touching spheres and the sphere;
Blue shading indicates the area from which the center of the third particle is excluded due to steric constraints.
(c,d) Minimum-energy membrane shapes for three spheres  at $\theta = 0\degree$ in (c) and $\theta = 90\degree$ in (d) for distance $\ell = 5$ $\upmu$m and confinement volume  $v = 7$ $\upmu{\rm m}^3$ per sphere. 
(e,f) Minimum bending energy $G$  versus distance $\ell$ for three spheres  at $\theta = 0\degree$ in (e) and $\theta = 90\degree$ in (f) for confinement volume $v = 6$ (blue), $7$ (yellow), $8$ (green), and $9$ $\upmu{\rm m}^3$ (red) per sphere. Lines represent 12th order polynomial fits of minimization results (points). Insets show contour plots of the minimum-energy shapes in (c) and (d).
(g,h) Comparison of measured (data points) and calculated forces (lines) versus particle distance. The force curves correspond to derivatives of the energy curves in (e) and (f) for the membrane bending rigidities $\kappa = 18$ $k_BT$ (dashed lines) and $22$ $k_BT$ (full lines). The data points represent mean values with error of the mean at distance $\ell$ for $\theta = 0\degree$ to $6\degree$ in (g) and $\theta = 84\degree$ to $90\degree$ in (h).
}

\end{figure*}

With the interaction force between two spheres known, we now turn to three spherical particles to start investigating many-body effects. To determine the preferred configuration of three particles, we initially arranged these particles in a line at a distance of about 3 times their diameter with optical tweezers (Fig.~\ref{fig:three_particles}a, $t = 0$ s). After release from the tweezers, the particles first move inwards while maintaining the linear arrangement before quickly rearranging into a close-packed equilateral triangle at $t = 4$ s (see Fig.~\ref{fig:three_particles}a and SI Movie~S3). Once formed, this particle triangle remained stable. 

We measure the force experienced by a third particle in the presence of two touching spheres (a ``dumbbell'') by holding all particles in separate optical traps (SI Movie~S4). 
To do so, we introduce two parameters: $\ell$, which is the distance from the center of the third particle to the center of mass of the particle dumbbell, and $\theta$, which is the angle between the long axis of the dumbbell and the center of the particle, see Figure~\ref{fig:three_particles}b inset. We find that the particle is always attracted towards the pair of particles, as shown in the polar plot of the force as a function of $\ell$ (binned per 0.3~$\mu$m) and $\theta$ (binned per 15$^\circ$). Arrows indicate both the direction and magnitude of the force, see Figure \ref{fig:three_particles}b. 
At larger distances of $\ell > 2.5\mu$m, the force along the long and the short axis of the pair is the same within error bounds and continuously increases with shorter distances. Notably, compared to the two-body force, the magnitude of the force at close distances did not increase but was rather reduced, which is a fingerprint of non-additivity. 

 We again compare our experimental results to numerical calculations by determining the minimum-energy shapes of three particles in linear orientation, i.e. with $\theta = 0\degree$ (Fig.~\ref{fig:three_particles}c), and triangular orientation, i.e. with $\theta=90\degree$ (Fig.~\ref{fig:three_particles}d). At large distances $\ell$, the bending energy $G$ is identical for the two orientations at the same values of $v$ (Fig.~\ref{fig:three_particles}e,f). At particle contact, which occurs at a distance $\ell = 1.88$ $\upmu$m for $\theta = 0\degree$ and $\ell = 1.08$ $\upmu$m for $\theta = 90\degree$, however, the bending energy of the triangular conformation is significantly lower. For example, at the confinement volume $v = 7$ $\upmu{\rm m}^3$ per particle, the bending energy is lower by ~$0.69\,\kappa \approx 14\,k_BT$ for $\kappa\approx 20\,k_BT$  which explains the experimental observation of Fig.~\ref{fig:three_particles}a. The calculated force profiles determined from derivatives of the bending energy profiles are in overall good agreement with the measured forces (Fig.~\ref{fig:three_particles}g,h).

\begin{figure}[hbt!]
\centering
\includegraphics[width=8.7cm]{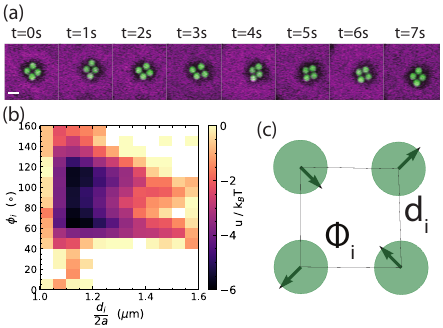}
\caption{\label{fig:four_particles}\textbf{Membrane-mediated interactions between four spheres}
(a) Snapshots of the time series evolution of a cluster consisting of four spheres underneath a vesicle with $h_v=22$~$\mu$m 
and $2R=31$~$\mu$m
; scale bar is 2~$\mu$m. 
(b) Interaction energy ($\frac{u}{k_BT}$) as a function of $d_i/2a$ and $\phi_i$.
(c) Schematic illustration of the most prominent mode for four particles (arrows) and indication of parameters to describe the quadrilateral configuration, with sides of $d_i$ and inner angles of $\phi_i$.
}
\end{figure}

Having established from the measurements of two and three particles that the interaction is strong, non-additive and that more compact arrangements seem to be favored, we now turn to four membrane-deforming particles. Bringing four spheres underneath a vesicle, we observe a quick formation of a compact cluster. The four particles form a square that dynamically rearranges into the two opposing diamond configurations, see Figure \ref{fig:four_particles}a and SI Movie~S5. We quantify the arrangements by measuring the distances $d_i$ and interior angles $\Phi_i$ between all particles, where $i$ labels the particle number.  We calculated the free energy from the probability of finding $\Phi_i$ and the side lengths of the quadrilateral normalized by the particle diameter, $d_i/2a$, see Figure \ref{fig:four_particles} b and c. The minima of the interaction potential were found at 65$^\circ$ and 115$^\circ$ and $d/2a$ between 1.1 and 1.2, corresponding to the diamond configurations. The minima have a depth of -6~$k_BT$ and the differences between the diamond and the square configuration are less than 1${\rm k}_B$T, allowing for easy reconfiguration, see SI Figure S2 and movie S5. These results are in agreement with our minimization approach, where the energy difference between the diamond and square configurations is found to be less than 0.3 $k_BT$ for all considered values of the confinement volume $v$ per particle between 6 and 9 $\upmu{\rm m}^3$. 

To better understand the reconfiguration of the four-body cluster, we measure the modes occurring in the structure and the stiffness associated with them by quantifying the eigenvalues and eigenvectors of the covariance matrix~\cite{Melio2024}. The most prominent mode is shown in Fig.~\ref{fig:four_particles}c, which represents a diagonal motion of two opposing particles towards or away from each other, i.e.~a reconfiguration between square and diamond arrangements. A projection of the particle position onto this mode reveals a double peak each with stiffness 30~nN/m each, which needs only 30~fN to move 1~$\upmu$m towards this mode. All modes and the corresponding stiffnesses are shown in Fig.~S3. 
The four-particle cluster does not assume non-compact configurations such as a line as they would be energetically unfavorable, but can dynamically rearrange between different compact arrangements.

\begin{figure*}
\centering
\includegraphics[width=1\linewidth]{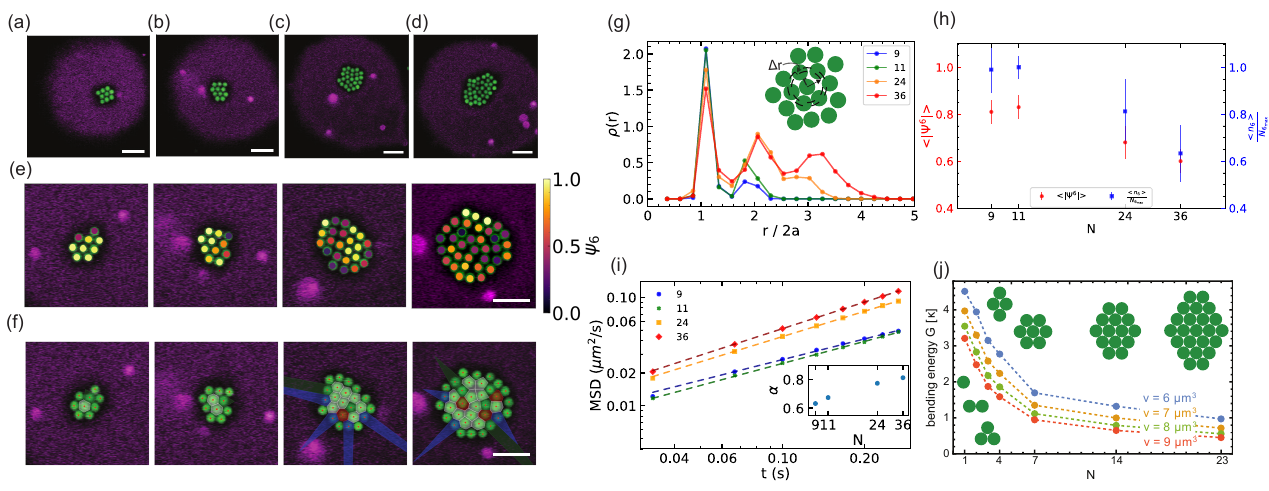}
\caption{\label{fig:many_body}\textbf{Membrane-mediated interactions between many spheres} 
Confocal microscopy images of (a) nine, (b) eleven, (c) twenty-four, and (d) thirty-six spheres confined underneath a vesicle ($h_v=34.0$~$\mu$m and $2R=49.8$~$\mu$m) show compact clusters. 
(e, f) From left to right (e) $\psi^6$ and (f) Voronoi diagram for nine, eleven, twenty-four and thirty-six spheres. Colors in (h) indicate number of neighbors: (5) yellow, (6) white, (7) blue, others shown in red. 
(g) Radial density profile $\rho(r)$ as a function of normalized distance $\frac{r}{2a}$ where $a$ is the particle radius. Scale bars are 5~$\mu$m. 
(h) Average order parameter $\langle|\psi ^6|\rangle$ (red data, left axis) and average number of particles with six neighbors $\langle n_6\rangle$ divided by the maximum possible number of particles with six neighbors $N_{6,max}$  (blue data, right axis) as function of number of particles $N$.
(i) Average mean square displacements (MSD) of  spheres in the clusters with N = 9, 11, 24, and 36 spheres. Inset shows the diffusion exponent $\alpha$. 
(j) Bending energy $G$ of the membrane around clusters of $N$ particles obtained from energy minimization at different values of the confinement volume $v$ per particle.
}

\end{figure*}

Our new model system allows us to straightforwardly study many-particle arrangements by simply dragging more spheres under a GUV. We show snapshots of the resulting clusters of $N$=9, 11, 24 and 36 particles in Fig. \ref{fig:many_body}a-d). For small numbers of particles, we see compact clusters with hexagonal order. The hexagonal order is in line with our observations of the triangle and diamond to be the preferred arrangements for three and four spheres as they are just a part of a hexagonal lattice. The compact shape on the other hand minimizes the overall bending energy of the membrane. 
For increasing numbers of particles, from $N=9$, 11, 24, and 36 particles, it becomes apparent that the resulting particle cluster remains compact, but has increasingly disordered arrangements as more particles are added (see Fig.~\ref{fig:many_body}a, b, c, and d). Hexagonal ordering has been observed in experiments with colloids electrostatically adsorbed onto surfactant vesicles~\cite{Ramos1999} and deformable microgel spheres attached to GUVs,~\cite{Mihut2013,Wang2019}, however, other interactions in addition to those stemming from membrane-deformations likely were present there as well. The triangular and linear arrangements observed for three fully membrane-wrapped spheres \cite{Azadbakht2023BPJ} also agree with hexagonal order, despite the absence of an adhesive energy in the present study.  

To quantify the order, we compute the radial density profile $\rho(r)$, which represents the number of particles within a certain radial distance $r$ from the center of the cluster.   The radial density profile is a parameter to probe crystalline order because it has very sharp peaks in an ordered solid and softens in the disordered liquid phase. It is defined as $\rho(r) = n(r)/ A(r)$ where $n(r)$ represents the number of particles found in a shell of thickness ${\rm d}r$ and area $A(r)$ at a distance $r$ from the center particle. As can be seen in Figure~\ref{fig:many_body}g, the radial density profiles $\rho(r)$ for $N=9$ and 11 spheres exhibit  sharp peaks for both the nearest neighbors at $r=1.25\mu$m as well as the next nearest neighbors. However, as the number of colloids increases, the peaks in $\rho(r)$ become wider, indicating a less structured liquid-like particle arrangement (see Fig.~\ref{fig:many_body}g and semi-log presentation in Fig.~S2). In addition, the peaks in the radial density profile become shifted to larger distances $r$ with increasing $N$. Measurements on non-fluorescently labelled colloids revealed that the membrane is still deformed by the colloids as shown in Fig.~S5, and we thus conclude that particles still interact by membrane-bending induced interactions, albeit at a possibly reduced strength, and with effective minima at larger distances.

To quantify the hexagonal order of the particle clusters, we use the local bond orientational order parameter $\Psi_6(r_i)=\frac{1}{N}\sum_j\exp[6i\theta(r_{ij})]$~\cite{Halperin1978}, which measures the average deviation of the angular orientation of neighboring membrane-deforming spheres from perfect hexagonal order, such that a value of 1 represents perfect hexagonal packing and a value of zero indicates disorder. In Fig.~\ref{fig:many_body}e and SI Movie~S6, S7, S8 and S9, it can be seen that the hexagonal order per particle decreases with increasing $N$, which can be quantified by the decrease of the averaged order parameter $\langle|\Psi_6|\rangle$ from about 0.8 for small particle numbers to about 0.6 for $N=36$, where the average is taken over all particles in the cluster and frames, see Fig.~\ref{fig:many_body}h. Similarly, the average number of particles with 6 nearest neighbors, $\langle n_6\rangle$, divided by the maximum possible number of particles with 6 nearest neighbors for a cluster of the same size, $N_{6,max}$, decreases from 1 to 0.63 with increasing particle number. While the number of particles with 6 neighbors attains the maximal value $\langle n_6\rangle/N_{6,max}= $ for clusters with  $N=9$ and 11 particles, the cluster composed of $N=24$ spheres on average had only 8.1~$\pm$~1.4 particles with 6 nearest neighbors while the maximum would have been ten particles with six neighbors (Movie~S10). For $N=36$ we measured an average of 12.0~$\pm$~2.3 particles with 6 nearest neighbors, compared to  $N_{6,max}=19$ (Movie~S11) see Figure~\ref{fig:many_body}~h. The decrease of $\langle n_6\rangle/N_{6,max}$ with increasing particle number of the clusters does not result from a non-spherical shape of the larger clusters, because the sphericity remains above 0.98 for all clusters, but is rather caused by the reduced order of the particles in the cluster.

The reduced degree of order in larger particle clusters leads to the appearance of defects. We identify them using a Voronoi tesselation of the surface and color the resulting tiles according to the number of neighbors, i.e.~4 neighbors are colored blue, 5 neighbors are colored green, 6 white, 7 red, and 8 purple. The cluster of 9 particles consistently had one central particle with six neighboring spheres, as seen in Movie~S12, and the cluster with eleven particles contained two spheres with six neighboring spheres, Movie~S13. The defects appearing for $N=9$ and $11$ only stem from the fact the clusters were one particle short to form a convex arrangement. In contrast, for $N=24$ and $36$, pairs of 5-7 fold defects appear on the inside of the cluster, see Fig.~\ref{fig:many_body}~f, in line with the reduced order identified by $\rho(r)$ and $\langle|\Psi_6|\rangle$. 

The increasing disorder in the particle arrangement with increasing $N$ is accompanied by a gradual increase in the diffusive motion of the particles. We quantify this by calculating the average mean squared displacement (MSD) of the particles in the cluster, see Fig.~\ref{fig:many_body}i, and find that the diffusion coefficient increases with $N$. We fit the obtained data with $\Delta r^2 = 4D \Delta t^\alpha$, where $\Delta r^2$ is the mean squared displacement, $D$ is the diffusion coefficient, and $\alpha$ is the diffusion exponent, and find that $\alpha$ is smaller that 1, which implies that the confinement of the particles induces subdiffusive motion. With increasing $N$, $\alpha$ tends to 1, i.e.\ the particle's motion approaches the diffusive regime, as shown in the inset of Fig.~\ref{fig:many_body}i. 

To understand the origin of these observations we determine the minimum energy shapes of the membrane around densely-packed, regular clusters of 7, 14, and 23 particles for different values of the confinement volume $v$ per particle for comparison. Fig.\ \ref{fig:many_body}j shows the overall bending energy $G(N)$ of the membrane for these particle clusters, as well as for a single particle, two particles in contact, three contacting particles in triangular conformation, and four particles in diamond conformation. We find that $G(N)$ strongly decreases with increasing number $N$ of particles in the cluster. For a confinement volume $v=7$  $\upmu{\rm m}^3$ per particle, for example, the bending energy of the membrane around a single particle is $G(1)=3.97\;\kappa$, while the membrane bending energy for two particles in contact decreases to $G(2) = 3.30\;\kappa$, and the bending energy $G(N)$ of the membrane around clusters of $N =3$, 4, 7, 14, and 23 particles is 2.58, 2.24, 1.35, 1.00, and 0.72 $\kappa$, respectively. We attribute this strong decrease in membrane bending energy with $N$ to the many-body effects in membrane-bending induced interactions. The solvent pockets around the particles play a crucial role here as their coalescence allows for longer-ranged, energetically significantly less costly membrane deformations around larger particle clusters. The increased disorder of the larger particle clusters observed in our experiments can be understood from the fact that there are more particles that can take up thermal energy, while the overall membrane bending energy that holds the increased particle cluster together is reduced, which combined leads to a stronger role of entropic particle repulsion in larger clusters.

To quantify the contribution of many-body effects we calculate the membrane bending energy that is released upon bringing $N$ single particles from a large distance where they do not interact into contact, which is $G(N) - NG(1)$, listed in the first line of Table 1. 
Clearly, this total interaction energy is dominated by the term $-NG(1)$ for large $N$, because $G(N)$ strongly decreases with $N$. We then determine the sum of the two-body interactions in the particle clusters using the two-particle interaction energy profiles of Fig.\ 1f. For $N=3$ particles, for example, the overall two-body interaction energy of $-13.9\; \kappa$ (see Table 1) is three times the two-particle interaction energy $-4.64\;\kappa$, because all three particles are in contact. The many-body interaction of the three particles, i.e.\ the three-body interaction, then simply is the difference between the total interaction energy and the sum of the two-body interactions. This three-particle interaction of $4.6\; \kappa$ is positive, which reflects cooperativity effects leading to total interaction energies that are less than the sum of the two-body interactions. With increasing $N$, the overall many-body interactions in the clusters increase in magnitude due to a larger and larger mismatch between the total interaction energy and the sum of the two-body interactions.

\begin{table}
\centering
\caption{Interaction energies in particle clusters for $v = 7$ $\upmu{\rm m}^3$: The first row lists the total  membrane bending energy for a cluster of size $N$ released when single particles are brought in contact. The second row lists the interaction energy expected from summing up two-body interactions. The third row lists the difference between row 1 and row 2 which can be attributed to many-body effects. All energies are given in units of $\kappa$.}
\begin{tabular}{l|cccccc}
$N$ particles & 2 & 3 & 4 & 7 & 14 & 23 \\
\hline
total & $-4.64$ & $-9.3$ & $-13.6$ & $-26.4$ & $-54.6$ & $-90.6$ \\
two-body & $-4.64$ & $-13.9$ & $-26.2$ & $-80.6$ & $-240.1$ & $-462.3$ \\
many-body & $0$ & $4.6$ & $12.6$ & $54.2$ & $185.5$ & $371.7$ \\
\bottomrule
\end{tabular}

\end{table}

\section*{Conclusion}
We have designed and exploited a simple model system of spherical colloids pulled underneath sessile GUVs to quantify the membrane deformation-mediated interactions of spherical, symmetric particles. We quantified the force and interaction energy between two and three particles and found that due to the large deformation caused by the particles the attraction extends over several particle diameters and is of the order of 100 $k_B T$ at close particle distances. We found that the non-additivity strongly reduces the attraction upon addition of particles. For small particle numbers, we observe hexagonally close-packed arrangements of compact, spherical clusters, pointing at the dominant influence of membrane bending energy. 
With increasing number of particles, the hexagonal order unexpectedly decreased due to a combination of a repulsive many-body effect and the additional solvent volume accessible to the particles. While maintaining a compact circular cluster shape at all times, the particles can rearrange between multiple conformations and with increasing number of particles, defects appear and the diffusion coefficient increases.

Our powerful model system does not require adhesive interactions between particles and membranes, and hence can be straightforwardly extended to study the membrane-mediated interactions between anisotropic objects, which more closely mimic the complex shapes of proteins and can be realized for example by direct laser writing of colloids~\cite{Doherty2020}. In addition, deformable colloids or vesicles could be used to identify the influence of elasticity.\cite{Midya2022} The addition of an adhesive force between membrane and particles would be a way to break the membrane-bending induced confinement into circular arrangements and allow us to test predictions about linear aggregates and membrane tubulation.

\matmethods{
\subsection*{Chemical}

D-glucose, sucrose, chloroform (\ch{CHCl3}), acetic acid, \ch{KOH}, poly(acrylamide) solution, EDC  N-(3-Dimethylaminopropyl)-N'-ethylcarbodiimide hydrochloride (98\%),
Sulfo-NHS ( N-Hydroxysulfosuccinimide sodium salt), N,N,N$^\prime$,N$^\prime$-Tetramethyl ethylenediamine (TEMED), and ammonium persulfate (APS),3-(trimethoxysilyl)propyl methacrylate (TPM) were purchased from Sigma-Aldrich. 
DOPE-rhodamine  (1,2-dioleoyl-sn-glycero-3-phosphoethanolamine-N-(lissaminerhodamine B sulfonyl)), DOPC ( $\Delta$ 9-cis 1,2-dioleoyl-sn-glycero-3-phosphocholine),
DOPE-PEG2000 (1,2-dioleoyl-sn-glycero-3-phosphoethanolamine-N-[(polyethylene glycol)-2000]) were provided from Avanti Polar Lipids. 
Sodium azide  (\ch{NaN3}) 99\% extra pure, and Potassium chloride (\ch{KCl}) 99+\%,
was obtained from Acros Organics, 
phosphate buffered saline (PBS) tablets from Merck Millipore, ethanol (\ch{C2H5OH}) from VWR. mPEG5000-NH$_2$  Methoxypolyethylene glycol amine M.W. 5000 bought Alfa Aesar. All water (\ch{H2O}) used was filtered with a MilliQ MilliPore apparatus (resistivity 18.2 M$\Omega\cdot$ cm). All chemicals were used as received.

\subsection*{Vesicle Preparation}
DOPC, DOPE-PEG2000-biotin, and DOPE-Rhodamine were mixed in chloroform at concentrations of 97.5\%wt, 2\%wt, and 0.5\%wt, respectively. The lipid mixture was dried on indium tin oxide (ITO) electrodes in a vacuum for at least 2h. Afterwards, giant vesicles were obtained from the dried lipids by electroformation by applying an alternating current at 1.1 $V_{rms}$ and 10 Hz for 2h.

\subsection*{Coverslip Functionalization}
We functionalized the coverslips to prevent adhesion of GUVs and particles by first coating the coverslips with TPM before polymerizing a layer of acrylamide on top, as described in ~\cite{Lau2009}. In summary, coverglasses were first cleaned by sonication in 1M KOH  solution, followed by washing once with ethanol and three times with MiliQ water. Then, functionalization with is achieved by submerging the coverslides in ethanol containing 1\%v/v acetic acid and 0.5\%v/v TPM for 15 minutes, before washing the slides three times with ethanol, and incubating them for one hour at 80 °C. Subsequently, polymerization  of acrylamide is carried out for 2 hours in a 2\%w/w acrylamide solution (evacuated in vacuum for 30 minutes to remove oxygen), with 0.035\%v/w TEMED and 0.070\%w/w APS. The resulting cover glasses were stored in a fridge immersed in the polymerization solution. We rinsed the coverglasses with water before using it.

\subsection*{Colloidal Particle Preparation}
Polystyrene (PS) particles were prepared by surfactant-free radical polymerization yielding spheres $1.25 \pm 0.05 \mu$m in diameter and highly carboxylated surface~\cite{Appel2013}. The fluorescent dye BODIPY was also added during synthesis for imaging. Subsequently, PS particles were functionalized with mPEG5000 according to the procedure described in \cite{VanDerWel2016,VanDerWel2017a}.

\subsection*{Microscopy and Optical Trapping}
Images were acquired using an A1R Nikon confocal scanner on a Ti-E microscope with a 60X water immersion objective (N.A. = 1.2). To increase the scanning speed, the confocal scanner was operated in resonance mode, resulting in 29 frames/s  512 $\times$ 512 pixels in size, at a resolution of 102 px/nm.
Particles were stained with BODYPI excited with a 488nm laser, and emission light was collected at 500-500 nm. Vesicles were doped with 0.5 \% rhodamine which was excited with a 561 nm laser, and the emission was collected in the range of 570-620 nm. Two separate photodetectors were used to simultaneously detect two fluorophores.
A 1064 nm laser beam from LaserQUANTUM was expanded to fill the aperture of a Meadowlark Spatial Light Modulator (SLM) (HS1920). The SLM modulates the phase of the laser wavefront and is imaged onto the back focal plane of the microscope objective through two plano-convex lenses. In the front focal point of the first lens, the non-diffracted light was filtered out. The light then is directed into the light path of the microscope via a dichroic mirror to allow simultaneous imaging and trapping. 
The SLM has been programmed by RedTweezers software\cite{Bowman2014a} run on a Geforce RTX4000 GPU. This setup generates 120 holograms per second while the SLM refreshing speed is 120 frames/sec. 

\subsection*{Force Measurement}
The force exerted on a trapped particle is given by 
\begin{equation}\label{eq:force}
F=k_{trap} (x-x_0)    
\end{equation}
where $x$ is the position of the particle and $x_0$ denotes the equilibrium position and $k_{trap}$ is the stiffness of the trap. This force can be acquired in both $x$ and $y$ directions. Optical traps can be easily calibrated by the equipartition theorem to quantify $k_{trap}$. Here we used the position variance $\sigma_x^2$ to estimate it by 
$k_{trap}= \frac{k_BT}{\sigma_x^2}$
where $k_B$ is Boltzmann constant and $T$ is the experimental temperature. 

\subsection*{Energy Minimization} We describe the membrane shape in Monge parametrization by the height $h(x,y)$ above a reference $x$-$y$ plane, i.e.\ the midplane of the flat membrane on the substrate away from the particle positions. 
In this parametrization, the mean curvature of the membrane shape can be expressed as
\begin{equation}
H = \frac{(1 + h_{x}^2) h_{yy} + (1 + h_{y}^2) h_{xx} - 2 h_{x} h_{y} h_{xy}}{2 (1 + h_{x}^2 + h_{y}^2 )^{3/2}}
\end{equation}
with subscripts $x$ and $y$ indicating partial derivatives to determine the membrane bending energy 
\begin{equation}
G = \int 2 \kappa H^2 {\rm d}A
\end{equation}
with bending rigidity $\kappa$ and membrane area element

\begin{equation}
{\rm d}A = \sqrt{1 + h_{x}^2 + h_{y}^2}\;{\rm d}x\,{\rm d}y
\end{equation}

In our numerical energy-minimization approach, we discretize the reference plane into a square lattice with lattice constant 100 nm and express the partial derivatives in the bending energy by standard multivariate finite differences. To constrain the confinement volume $v$ per particle, we use a pressure $p$ as Lagrange multiplier to adjust $v$, i.e.~we minimize the energy $E = G + p V$ with $V = n_p v =  \int h\; {\rm d}x\,{\rm d}y$ where $n_p$ is the number of particles, and determine $V$ and $G$ as functions of $p$ for interpolation to values of $p$ at which the desired value of $v$ is obtained. All minimizations are performed with the function FindMinimum of Mathematica 13 \cite{Mathematica}. In these minimizations, the position and spherical shape of the particles are taken into account by constraints on the height $h(x,y)$, i.e.~by lower bounds for $h$ at lattice sites located under the particles to prevent an overlap of the membrane with the particles.

}

\showmatmethods{} 

\acknow{A.A. and D.J.K. thank Yogesh Shelke, and Julio Melio for preparing PAA-coated coverslips, as well Rachel Doherty for synthesizing polystyrene colloidal particles. A.A. and D.J.K. are grateful to Casper van der Wel for providing open-source packages for particle tracking. T.R.W. thanks the Max Planck Society for funding.}

\showacknow{} 
\bibliography{references}

\begin{thebibliography}{10}

\bibitem{Goulian1993}
M Goulian, R Bruinsma, P Pincus, {Long-Range Forces in Heterogeneous Fluid
  Membranes}.
\newblock {\em\protect\JournalTitle{Europhysics Letters}} \textbf{23}, 155
  (1993).

\bibitem{VanDerWel2016}
C van~der Wel, et~al., {Lipid membrane-mediated attraction between curvature
  inducing objects}.
\newblock {\em\protect\JournalTitle{Scientific Reports}} \textbf{6}, 1--10
  (2016).

\bibitem{Sarfati2016}
R Sarfati, ER Dufresne, {Long-range attraction of particles adhered to lipid
  vesicles}.
\newblock {\em\protect\JournalTitle{Physical Review E}} \textbf{94}, 2--7
  (2016).

\bibitem{Saric2013}
A {\v{S}}ari{\'{c}}, A Cacciuto, {Self-assembly of nanoparticles adsorbed on
  fluid and elastic membranes}.
\newblock {\em\protect\JournalTitle{Soft Matter}} \textbf{9}, 6677--6695
  (2013).

\bibitem{Yolcu2014a}
C Yolcu, RC Haussman, M Deserno, {The Effective Field Theory approach towards
  membrane-mediated interactions between particles}.
\newblock {\em\protect\JournalTitle{Advances in Colloid and Interface Science}}
  \textbf{208}, 89--109 (2014).

\bibitem{Weikl2018}
TR Weikl, {Membrane-Mediated Cooperativity of Proteins}.
\newblock {\em\protect\JournalTitle{Annual Review of Physical Chemistry}}
  \textbf{69}, 521--539 (2018).

\bibitem{Galatola2023}
P Galatola, JB Fournier, Many-body interactions between curvature-inducing
  membrane inclusions with arbitrary cross-sections.
\newblock {\em\protect\JournalTitle{Soft Matter}} \textbf{19}, 6157--6167
  (2023).

\bibitem{Saric2012}
A {\v{S}}ari{\'{c}}, A Cacciuto, {Fluid membranes can drive linear aggregation
  of adsorbed spherical nanoparticles}.
\newblock {\em\protect\JournalTitle{Physical Review Letters}} \textbf{108},
  1--5 (2012).

\bibitem{Laradji2020}
M Laradji, PB Kumar, EJ Spangler, {Adhesion and Aggregation of Spherical
  Nanoparticles on Lipid Membranes}.
\newblock {\em\protect\JournalTitle{Chemistry and Physics of Lipids}}
  \textbf{233}, 104989 (2020).

\bibitem{Azadbakht2023BPJ}
A Azadbakht, B Meadowcroft, J Májek, A Šarić, DJ Kraft, Nonadditivity in
  interactions between three membrane-wrapped colloidal spheres.
\newblock {\em\protect\JournalTitle{Biophysical Journal}} \textbf{123},
  307--316 (2024).

\bibitem{Ramos1999}
L Ramos, TC Lubensky, N Dan, P Nelson, DA Weitz, {Surfactant-mediated
  two-dimensional crystallization of colloidal crystals}.
\newblock {\em\protect\JournalTitle{Science}} \textbf{286}, 2325--2328 (1999).

\bibitem{Koltover1999}
I Koltover, JO R{\"{a}}dler, CR Safinya, {Membrane mediated attraction and
  ordered aggregation of colloidal particles bound to giant phospholipid
  vesicles}.
\newblock {\em\protect\JournalTitle{Physical Review Letters}} \textbf{82},
  1991--1994 (1999).

\bibitem{Zhu2023}
Y Zhu, A Sharma, EJ Spangler, M Laradji, Non-close-packed hexagonal
  self-assembly of janus nanoparticles on planar membranes.
\newblock {\em\protect\JournalTitle{Soft Matter}} \textbf{19}, 7591--7601
  (2023).

\bibitem{Spanke2020}
HT Spanke, et~al., {Wrapping of Microparticles by Floppy Lipid Vesicles}.
\newblock {\em\protect\JournalTitle{Physical Review Letters}} \textbf{125},
  1--9 (2020).

\bibitem{Azadbakht2023NL}
A Azadbakht, B Meadowcroft, T Varkevisser, A Saric, DJ Kraft, {Wrapping
  pathways of anisotropic dumbbell particles by giant unilamellar vesicles}.
\newblock {\em\protect\JournalTitle{Nano Letters}} \textbf{23}, 4267--4273
  (2023).

\bibitem{VanDerWel2017a}
C van~der Wel, et~al., {Surfactant-free Colloidal Particles with Specific
  Binding Affinity}.
\newblock {\em\protect\JournalTitle{Langmuir}} \textbf{33}, 9803--9810 (2017).

\bibitem{Faizi2020}
HA Faizi, CJ Reeves, VN Georgiev, PM Vlahovska, R Dimova, {Fluctuation
  spectroscopy of giant unilamellar vesicles using confocal and phase contrast
  microscopy}.
\newblock {\em\protect\JournalTitle{Soft Matter}} \textbf{16}, 8996--9001
  (2020).

\bibitem{Nagle2013}
JF Nagle, Introductory lecture: basic quantities in model biomembranes.
\newblock {\em\protect\JournalTitle{Faraday Discuss.}} \textbf{161}, 11--29;
  discussion 113--50 (2013).

\bibitem{Weikl1998}
TR Weikl, MM Kozlov, W Helfrich, {Interaction of conical membrane inclusions:
  Effect of lateral tension}.
\newblock {\em\protect\JournalTitle{Physical Review E - Statistical Physics,
  Plasmas, Fluids, and Related Interdisciplinary Topics}} \textbf{57},
  6988--6995 (1998).

\bibitem{Simunovic2015b}
M Simunovic, GA Voth, {Membrane tension controls the assembly of
  curvature-generating proteins}.
\newblock {\em\protect\JournalTitle{Nature Communications}} \textbf{6} (2015).

\bibitem{Melio2024}
J Melio, SE Henkes, DJ Kraft, Soft and stiff normal modes in floppy colloidal
  square lattices.
\newblock {\em\protect\JournalTitle{Physical Review Letters}} \textbf{132},
  078202 (2024).

\bibitem{Mihut2013}
AM Mihut, AP Dabkowska, JJ Crassous, P Schurtenberger, T Nylander, {Tunable
  adsorption of soft colloids on model biomembranes}.
\newblock {\em\protect\JournalTitle{ACS Nano}} \textbf{7}, 10752--10763 (2013).

\bibitem{Wang2019}
M Wang, et~al., {Assembling responsive microgels at responsive lipid
  membranes}.
\newblock {\em\protect\JournalTitle{Proceedings of the National Academy of
  Sciences of the United States of America}} \textbf{116}, 5442--5450 (2019).

\bibitem{Halperin1978}
BI Halperin, DR Nelson, {Theory of Two-Dimensional Melting}.
\newblock {\em\protect\JournalTitle{Physical Review Letters}} \textbf{41}, 121
  (1978).

\bibitem{Doherty2020}
RP Doherty, et~al., {Catalytically propelled 3D printed colloidal
  microswimmers}.
\newblock {\em\protect\JournalTitle{Soft Matter}} \textbf{16}, 10463--10469
  (2020).

\bibitem{Midya2022}
J Midya, T Auth, G Gompper, {Membrane-Mediated Interactions Between
  Nonspherical Elastic Particles}.
\newblock {\em\protect\JournalTitle{ACS Nano}} \textbf{17}, 1935--1945 (2023).

\bibitem{Lau2009}
AW Lau, A Prasad, Z Dogic, {Condensation of isolated semi-flexible filaments
  driven by depletion interactions}.
\newblock {\em\protect\JournalTitle{Europhysics Letters}} \textbf{87}, 48006
  (2009).

\bibitem{Appel2013}
J Appel, S Akerboom, RG Fokkink, J Sprakel, {Facile one-step synthesis of
  monodisperse micron-sized latex particles with highly carboxylated surfaces}.
\newblock {\em\protect\JournalTitle{Macromolecular Rapid Communications}}
  \textbf{34}, 1284--1288 (2013).

\bibitem{Bowman2014a}
RW Bowman, et~al., {Red tweezers: Fast, customisable hologram generation for
  optical tweezers}.
\newblock {\em\protect\JournalTitle{Computer Physics Communications}}
  \textbf{185}, 268--273 (2014).

\bibitem{Mathematica}
WR Inc., Mathematica, {V}ersion 13.0 (year?) Champaign, IL, 2021.

\end{thebibliography}

\end{document}